\documentclass[twocolumn,showpacs,preprintnumbers,amsmath,amssymb, prb]{revtex4}

\usepackage{graphicx}
\usepackage{dcolumn}
\usepackage{bm}

\begin{document}


\title{Green's function theory for spin-$\frac{1}{2}$ ferromagnets with an easy-plane exchange anisotropy}

\author{Daisuke Yamamoto$^1$}
\email{yamamoto@kh.phys.waseda.ac.jp}
\author{Synge Todo$^2$}
\author{Susumu Kurihara$^1$}
\affiliation{$^1$Department of Physics, Waseda University, Okubo, Shinjuku-ku, Tokyo 169-8555, Japan }
\homepage{http://www.kh.phys.waseda.ac.jp/}
\affiliation{$^2$Department of Applied Physics, University of Tokyo, Hongo, Bunkyo-ku, Tokyo 113-8656, Japan }
\date{\today}

\begin{abstract}
The many-body Green's function theory with the random-phase approximation is applied to the study of easy-plane spin-1/2 ferromagnets in an in-plane magnetic field. We demonstrate that the usual procedure, in which only the three Green's functions $\langle\langle S_i^\mu;S_j^-\rangle\rangle $ ($\mu=+,-,z$) are used, yields unreasonable results in this case. Then the problem is discussed in more detail by considering all combinations of Green's functions. We can derive one more equation, which cannot be obtained by using only the set of the above three Green's functions, and point out that the two equations contradict each other if one demands that the identities of the spin operators are exactly satisfied. We discuss the cause of the contradiction and attempt to improve the method in a self consistent way. In our procedure, the effect of the anisotropy can be appropriately taken into account, and the results are in good agreement with the quantum Monte Carlo calculations. 
\end{abstract}

\maketitle

\section{\label{1}Introduction} 
Many-body Green's function theory is a powerful tool for theoretical studies of spin systems. In this formalism, a certain decoupling approximation is generally required to terminate the infinite hierarchy of equations of motion for higher-order Green's functions. The first (lowest) order decoupling scheme introduced by Tyablikov,\cite{Tyablikov} which is called the random-phase approximation (RPA) or ``Tyablikov decoupling,'' is a very simple yet effective way to perform this operation. Many previous authors have attempted to go beyond Tyablikov's method and generated a variety of decoupling procedures such as the Callen decoupling,\cite{Callen} Tahir-Kheli's theory,\cite{Tahir-Kheli} the modified versions of the Callen decoupling,\cite{Copeland,Swendsen} and Oguchi's variational theory.\cite{Oguchi} Nevertheless, it is known that the RPA decoupling is still the simplest and most reliable first order approximation [see, for example, the comparison between the RPA, the Callen decoupling, and the quantum Monte Carlo (QMC) calculations in Ref.~\onlinecite{Frobrich}]. More quantitative results may be obtained by applying the second-order Green's function theory originally proposed by Kondo and Yamaji~\cite{KY} for the one-dimensional isotropic Heisenberg model. In this theory, the hierarchy of equations of motion is terminated at the second step with introducing vertex parameters. Shimahara and Takada~\cite{ST} applied this theory to the two-dimensional case, and Junger {\it et al}.~\cite{Junger} and Antsygina {\it et al}.~\cite{Antsygina} extended it to the case where a uniform external field is applied. Several other methods~\cite{Frobrich,Rudoi12Wesselinowa,Reinecke} have also been attempted. However, some problems, e.g., how to determine the vertex parameters, are still under discussion, and thus these formalisms have not been fully established yet. Therefore, the simple RPA decoupling scheme is still often used for analytical studies of complicated systems.

Field-induced phenomena in spin systems have been the focus of both theoretical and experimental studies. For example, in uniaxially anisotropic Heisenberg antiferromagnets, the magnetic field applied along the easy-axis induces multicritical behavior at the triple point of the antiferromagnetic, spin-flop, and paramagnetic phases.\cite{Fisher,Rohrer,Landau,Ohgushi,Holtschneider} The spin reorientation transition induced by a transverse magnetic field has also attracted considerable interest for both ferromagnetic and antiferromagnetic cases.\cite{Frobrich,Reinecke,Rudoi4,Brown,Frobrich2,Frobrich3,Frobrich4,Jensen,Caux,Wang,Dmitriev,Schwieger,Pini,Jensen2,Yamamoto} Some of them were studied by using Green's function formalism with the RPA decoupling. However, in this paper, we point out that one should pay careful attention in applying the RPA decoupling scheme to such a complicated system, which has an anisotropy in the plane perpendicular to the direction of the magnetization. As an example of such systems, we consider spin-$1/2$ ferromagnets with an easy-plane exchange anisotropy. This system can be described by a comparatively simple model, and thus we can clarify the issues.

This paper is organized as follows. First, in Sec.~\ref{2}, we introduce the model Hamiltonian considered in this paper. In Sec.~\ref{3}, we demonstrate the application of the usual procedure, in which only a restricted set of Green's functions is used. In Sec.~\ref{4}, we show that one more equation, which cannot be obtained by the above procedure, can be derived by considering all combinations of Green's functions. Moreover, we point out that the two equations contradict each other if one demands that the identities of the spin operators are exactly satisfied. In Sec.~\ref{5}, the cause of the problems found in the previous section are discussed, and we attempt to improve the method in a self consistent way. The obtained results are compared with the QMC data. Finally, a summary is presented in Sec.~\ref{6}. 

\section{\label{2}Model Hamiltonian}
The many-body Green's function theory for spin systems, which is briefly reviewed in the Appendix, has been developed by many authors.\cite{Tyablikov,Callen,Tahir-Kheli,Copeland,Swendsen,Oguchi,Frobrich} We consider the application of this theory to the study of easy-plane ferromagnets in an in-plane magnetic field. The Hamiltonian of this system is given by
\begin{eqnarray}
H=-\frac{1}{2}\sum_{i,j}J_{ij}\!\left(S^x_i S^x_j\!+\!S^y_i S^y_j\!+\!\Delta S^z_i S^z_j \right)\!-\!h\!\sum_{i}S^x_i, \label{Hamiltonian}
\end{eqnarray}
and
\begin{eqnarray}
[S_i^\mu,S_j^\nu]=i\epsilon_{\mu\nu\lambda} S_i^\lambda\delta_{ij},\label{rule}
\end{eqnarray}
where $\mathbf{S}_{i}=(S_i^x,S_i^y,S_i^z)$ is the usual spin operator at site $i$ and $J_{ij}$ is the exchange interaction strength between site $i$ and site $j$. We take into account only the nearest-neighbour coupling, i.e., $J_{ij}=J>0$ if $i$ and $j$ are nearest-neighbour sites and $J_{ij}=0$ otherwise. In the spin-$1/2$ case, the single-ion anisotropy energy, e.g., $-\!\!\sum_{i}K_{2,i}(S_i^z)^2$, is constant and does not affect our results. The behavior of the system depends strongly on the value of the anisotropy parameter $\Delta$: for $0\leq\Delta<1$, the spins prefer to lie in the $xy$ plane (easy plane) while for $\Delta>1$, the spins tend to align along the $z$ axis (easy axis). In this paper, we focus on the easy-plane case.  The magnetic field $h$ in Eq.~(\ref{Hamiltonian}) is applied perpendicular to the hard ($z$) axis.

Let us calculate the magnetic properties of the model given by Eq.~(\ref{Hamiltonian}). Since the model is expressed in terms of spin operators ${\bf S}_i$, we have many choices of the operators $A$ and $B$ in Eq.~(\ref{Green}). First, the calculations with the usual set of Green's functions are shown in the next section. Then, considering all choices of Green's functions, we discuss the problem in more detail in the subsequent section. 
\section{\label{3}usual choice of Green's functions}
In the case of easy-plane anisotropy ($0\leq \Delta<1$), only the component of the magnetization parallel to the in-plane magnetic field has nonzero value, i.e, $\langle S_i^x \rangle=m$ and $\langle S_i^y \rangle=\langle S_i^z \rangle=0$.  In order to study the magnetic properties of spin-$S$ systems, the set of Green's functions $G_{ij,\eta}^{+-,mn}$, $G_{ij,\eta}^{--,mn}$, and $G_{ij,\eta}^{z-,mn}$ has been often used in many previous studies. Here we denote
\begin{eqnarray}
G_{ij,\eta}^{\mu-,mn}&=&\langle \langle S_i^\mu;(S_j^z)^m(S_j^-)^n\rangle\rangle_\eta,
\end{eqnarray}
where $m\geq 0$ and $n\geq 1$ are positive integers. $S_i^{\pm}=S_i^x \pm iS_i^y$ are the usual spin raising and lowering operators, and $\eta=-1$ ($\eta=+1$) denotes the retarded commutator (anticommutator) Green's function (see Appendix for the detailed definition of Green's function). The latter two ($G_{ij,\eta}^{--,mn}$ and $G_{ij,\eta}^{z-,mn}$) are used for relatively complicated systems, as needed.\cite{Frobrich,Frobrich2,Frobrich3,Frobrich4,Jensen,Wang} In this section, according to the previous works, we employ this usual set of Green's functions.

To avoid extra complexity, we consider the case of $S=1/2$, where the above usual set is reduced to only three Green's functions $G_{ij,\eta}^{+-}=\langle \langle S_i^+;S_j^-\rangle\rangle_\eta$, $G_{ij,\eta}^{--}=\langle \langle S_i^-;S_j^-\rangle\rangle_\eta$, and $G_{ij,\eta}^{z-}=\langle\langle S_{i}^{z};S_{j}^-\rangle\rangle_\eta$. The equations of motion for these Green's functions are given by
\begin{eqnarray}
\omega G_{ij,\eta}^{\mu -}(\omega)=\langle [S_i^\mu, S_j^-]_\eta\rangle+\langle\langle[S_{i}^{\mu},H];S_{j}^- \rangle\rangle_{\eta,\omega}, 
\label{Eqofmotion}
\end{eqnarray}
with
\begin{eqnarray}
[S_{i}^{\pm},H] = \mp \sum_k J_{ik}(S_i^z S_k^\pm - \Delta S_i^\pm S_k^z )\mp hS_i^z, 
\end{eqnarray}
and
\begin{eqnarray}
[S_{i}^{z},H] = -\frac{1}{2}\sum_k J_{ik}(S_i^+ S_k^- - S_i^- S_k^+ )\!-\!\frac{h}{2}(S_i^+ \!-\!S_i^-). 
\end{eqnarray}
The right-hand sides of the above equations include higher-order Green's functions, e.g., $\langle\langle S_i^zS_k^+;S_{j}^- \rangle\rangle_{\eta,\omega}$. In order to close the system of equations, we adopt the generalized version of the so-called RPA or Tyablikov decoupling,\cite{Tyablikov} 
\begin{eqnarray}
\langle\langle S_{i}^{\mu}S_{k}^{\nu};S_{j}^\lambda \rangle\rangle_{\eta,\omega}\!\!&\approx&\!\!\langle S_{i}^{\mu}\rangle \langle\langle S_{k}^{\nu};S_{j}^\lambda \rangle\rangle_{\eta,\omega}\!+\!\langle S_{k}^{\nu}\rangle \langle\langle S_{i}^{\mu};S_{j}^\lambda \rangle\rangle_{\eta,\omega}\nonumber\\
\!\!&&\!\!-\langle\langle\langle S_{i}^{\mu}\rangle \langle S_{k}^{\nu}\rangle;S_{j}^\lambda \rangle\rangle_{\eta,\omega}. 
\label{RPA}
\end{eqnarray}
Performing the Fourier transformations given by
\begin{eqnarray}
G_{{\bf q},\eta}^{\mu \nu}=\frac{1}{N}\sum_{i,j}G_{ij,\eta}^{\mu \nu}e^{-i{\bf q}\cdot({\bf R}_{i}-{\bf R}_{j})}, \label{Fourier}
\end{eqnarray}
and
\begin{eqnarray}
\Lambda_{{\bf q},\eta}^{\mu \nu}=\frac{1}{N}\sum_{i,j}\langle [S_i^\mu, S_j^\nu]_\eta\rangle e^{-i{\bf q}\cdot({\bf R}_{i}-{\bf R}_{j})}, 
\label{Fourier2}
\end{eqnarray}
where $N$ is the number of lattice sites, we now rewrite Eq.~(\ref{Eqofmotion}) in a matrix form: 
\begin{eqnarray}
\left[\omega {\bf 1}\!-\!\left(\!
 \begin{array}{ccc}
 0 & 0&-\tilde{\Gamma}_{{\bf q}}\\
 0 & 0&~~\tilde{\Gamma}_{{\bf q}}\\
 -\Gamma_{{\bf q}}/2&\Gamma_{{\bf q}}/2&0
 \end{array}\!
\right)\!\! \right]\!\!
\left(\!
 \begin{array}{c}
 G_{{\bf q},\eta}^{+ -}\\
 G_{{\bf q},\eta}^{- -}\\
 G_{{\bf q},\eta}^{z -}
 \end{array}\!
\right)\!\! =\!\!\left(\!
 \begin{array}{c}
\Lambda_{{\bf q},\eta}^{+ -}\\
\Lambda_{{\bf q},\eta}^{- -}\\
\Lambda_{{\bf q},\eta}^{z -}
 \end{array}\!
\right)
\label{mateq}
\end{eqnarray}
with $\tilde{\Gamma}_{{\bf q}}=h+zJm(1-\Delta\gamma_{\bf q})$ and $\Gamma_{{\bf q}}=h+zJm(1-\gamma_{\bf q})$, where $z=2$ ($z=4$) is the number of nearest neighbors and $\gamma_{\bf q}=\cos q_{x}$ [$\gamma_{\bf q}=\frac{1}{2}(\cos q_{x}+\cos q_{y})$] is the Fourier factor for a chain (for a square lattice). Here we set the lattice constant to be unity. 
By solving Eq.~(\ref{mateq}), one derives the commutator Green's functions
\begin{eqnarray}
G_{{\bf q},-1}^{\pm -}=\frac{\pm m \tilde{\Gamma}_{{\bf q}}}{\omega^2-\omega_{\bf q}^2},~~~G_{{\bf q},-1}^{z -}=\frac{-m \omega}{\omega^2-\omega_{\bf q}^2}
\label{Greens_function1}
\end{eqnarray}
with $\omega_{\bf q}^2=\tilde{\Gamma}_{{\bf q}}\Gamma_{{\bf q}}$. The anticommutator Green's functions $G_{{\bf q},+1}^{\pm -}$ have a pole at $\omega=0$ and one obtains
\begin{eqnarray}
C_{{\bf q},+1}^{\pm -}=\lim_{\omega\rightarrow 0}\omega G_{{\bf q},+1}^{\pm -}=\frac{\Lambda_{{\bf q},+1}^{+ -}+\Lambda_{{\bf q},+1}^{- -}}{2}.\label{CP1}
\end{eqnarray}
Then the spectral theorem [Eq.~(\ref{spectral3})] gives 
\begin{eqnarray}
\frac{\Lambda_{{\bf q},+1}^{+ -}-\Lambda_{{\bf q},+1}^{- -}}{2}=\frac{m \tilde{\Gamma}_{{\bf q}}}{\omega_{\bf q}}\coth \frac{\beta \omega_{\bf q}}{2}, \label{mag01}
\end{eqnarray}
and
\begin{eqnarray}
\Lambda_{{\bf q},+1}^{z -}=0. \label{mag02}
\end{eqnarray}
Note that the applications of the spectral theorem to $G_{{\bf q},\eta}^{+ -}$ and $G_{{\bf q},\eta}^{- -}$ yield the same equation [Eq.~(\ref{mag01})]. Moreover, the following expression is derived from Eq.~(\ref{mag01}): 
\begin{eqnarray}
\frac{\langle\{S_i^+, S_i^-\}\rangle}{2}-\langle (S_i^-)^2\rangle=\frac{1}{N} \sum_{\bf q}\frac{m \tilde{\Gamma}_{{\bf q}}}{\omega_{\bf q}}\coth \frac{\beta \omega_{\bf q}}{2}. \label{mag012}
\end{eqnarray}
Here the sum runs over the first Brillouin zone. In addition to the relation for general spins $S$, 
\begin{eqnarray}
\{S_i^+ ,S_i^-\}=2\left[S(S+1)-(S_i^z)^2\right], \label{sum_rule}
\end{eqnarray}
the spin-$1/2$ operators have the following properties: 
\begin{eqnarray}
(S_i^+)^2=(S_i^-)^2=0,~~(S_i^z)^2=\frac{1}{4}. \label{kinematic_interaction}
\end{eqnarray}
If, as usual, we demand that the conditions given by Eqs.~(\ref{sum_rule}) and (\ref{kinematic_interaction}) are satisfied even within the RPA, Eq.~(\ref{mag012}) becomes
\begin{eqnarray}
\frac{1}{2}=\frac{1}{N} \sum_{\bf q}\frac{m \tilde{\Gamma}_{{\bf q}}}{\omega_{\bf q}}\coth \frac{\beta \omega_{\bf q}}{2}\label{mag03}. 
\end{eqnarray}
One can now calculate the magnetization $m$ from this self-consistent equation. 

The above procedure has been often used in the case of systems with an easy-axis anisotropy ($\Delta>1$) and a transverse magnetic field.\cite{note} Despite this, the obtained results are in poor agreement with the numerical results as we shall discuss later. 

\section{\label{4}Contradiction between two equations}
In the previous section, according to some previous studies, we use the set of Green's functions $\{G_{ij,\eta}^{+-}, G_{ij,\eta}^{--}, G_{ij,\eta}^{z-}\}$ and adopt the conditions described by Eqs.~(\ref{sum_rule}) and (\ref{kinematic_interaction}). However, this choice of Green's functions is not enough to take care of all directions in spin space. In this section, we consider all combinations of spin operators on the choice of the operators $A$ and $B$ in Eq.~(\ref{Green}): 
\begin{eqnarray}
G_{ij,\eta}^{\mu \nu}=\langle \langle S_i^\mu;S_j^\nu\rangle\rangle_\eta ~~~(\mu,\nu=x,y,z). 
\label{Green2}
\end{eqnarray}
The equations of motion are given by
\begin{eqnarray}
\omega G_{ij,\eta}^{\mu \nu}(\omega)=\langle [S_i^\mu, S_j^\nu]_\eta\rangle+\langle\langle[S_{i}^{\mu},H];S_{j}^\nu \rangle\rangle_{\eta,\omega}, 
\label{Eqofmotion2}
\end{eqnarray}
with
\begin{eqnarray}
[S_{i}^{x},H] \!\!&=&\!\! -i\sum_k J_{ik}(S_i^z S_k^y - \Delta S_i^y S_k^z ), \label{dtSx}\\
\ [S_{i}^{y},H] \!\!&=&\!\! -i\sum_k J_{ik}(\Delta S_i^x S_k^z - S_i^z S_k^x )+i hS_i^z, \label{dtSy}
\end{eqnarray}
and
\begin{eqnarray}
[S_{i}^{z},H] = -i\sum_k J_{ik}(S_i^y S_k^x - S_i^x S_k^y )-i hS_i^y. \label{dtSz}
\end{eqnarray}
Applying the RPA decoupling expressed by Eq.~(\ref{RPA}), and performing the Fourier transformations given by Eqs.~(\ref{Fourier}) and (\ref{Fourier2}), one derives the following matrix equation instead of Eq.~(\ref{mateq}): 
\begin{eqnarray}
\left[\omega {\bf 1}-\left(
 \begin{array}{ccc}
 0&0&0\\
 0&0 & i\tilde{\Gamma}_{{\bf q}}\\
 0&-i\Gamma_{{\bf q}} & 0
 \end{array}
\right) \!\right]\!
\left(
 \begin{array}{c}
 G_{{\bf q},\eta}^{x \nu}\\
 G_{{\bf q},\eta}^{y \nu}\\
 G_{{\bf q},\eta}^{z \nu}
 \end{array}
\right)\! =\!\left(
 \begin{array}{c}
\Lambda_{{\bf q},\eta}^{x \nu}\\
\Lambda_{{\bf q},\eta}^{y \nu}\\
\Lambda_{{\bf q},\eta}^{z \nu}
 \end{array}
\right). 
\label{mateq2}
\end{eqnarray}
By solving Eq.~(\ref{mateq2}), one obtains the commutator Green's functions
\begin{eqnarray}
G_{{\bf q},-1}^{x \nu}&=&0,\label{Greens_functionx}\\
G_{{\bf q},-1}^{y \nu}&=&\frac{\omega\Lambda_{-1}^{y \nu}+i \tilde{\Gamma}_{{\bf q}}\Lambda_{-1}^{z \nu}}{\omega^2-\omega_{\bf q}^2},\label{Greens_functiony}
\end{eqnarray}
and
\begin{eqnarray}
~~G_{{\bf q},-1}^{z \nu}=\frac{\omega\Lambda_{-1}^{z \nu}-i \Gamma_{{\bf q}}\Lambda_{-1}^{y \nu}}{\omega^2-\omega_{\bf q}^2}.
\label{Greens_functionz}
\end{eqnarray}
Since the commutator of spin operators is local in site indices, $\Lambda_{{\bf q},-1}^{\mu \nu}$ are independent of ${\bf q}$ and thus we drop the subscript ${\bf q}$ from $\Lambda_{{\bf q},-1}^{\mu \nu}$. Here we use the fact $\Lambda_{-1}^{x \nu}=0$, which is valid for any $\nu$. Additionally, one obtains
\begin{eqnarray}
C_{{\bf q},+1}^{x \nu}=\Lambda_{{\bf q},+1}^{x \nu},~~C_{{\bf q},+1}^{y \nu}=C_{{\bf q},+1}^{z \nu}=0 \label{CP2}
\end{eqnarray}
from the anticommutator Green's functions. 
Then the spectral theorem [Eq.~(\ref{spectral3})] gives 
\begin{eqnarray}
\Lambda_{{\bf q},+1}^{y \nu}=\frac{i\tilde{\Gamma}_{{\bf q}}\Lambda_{-1}^{z \nu}}{\omega_{\bf q}}\coth \frac{\beta \omega_{\bf q}}{2}, \label{mag1}
\end{eqnarray}
and
\begin{eqnarray}
\Lambda_{{\bf q},+1}^{z \nu}=\frac{-i\Gamma_{{\bf q}}\Lambda_{-1}^{y \nu}}{\omega_{\bf q}}\coth \frac{\beta \omega_{\bf q}}{2}. \label{mag2}
\end{eqnarray}
Incidentally, the application of the spectral theorem to $G_{{\bf q},\eta}^{x \nu}$ gives the trivial equation $\Lambda_{{\bf q},+1}^{x \nu}=\Lambda_{{\bf q},+1}^{x \nu}$. Then Eqs.~(\ref{mag1}) and (\ref{mag2}) become, for $\nu=y$ and $\nu=z$, respectively, 
\begin{eqnarray}
\langle (S_i^y)^2 \rangle=\frac{1}{N} \sum_{\bf q}\frac{m \tilde{\Gamma}_{{\bf q}}}{2\omega_{\bf q}}\coth \frac{\beta \omega_{\bf q}}{2},\label{mag3}
\end{eqnarray}
and
\begin{eqnarray}
\langle (S_i^z)^2 \rangle=\frac{1}{N} \sum_{\bf q}\frac{m \Gamma_{{\bf q}}}{2\omega_{\bf q}}\coth \frac{\beta \omega_{\bf q}}{2}. \label{mag4}
\end{eqnarray}
If we demand that the property specific to the spin-$1/2$ operators, $(S_i^y)^2=1/4$, is satisfied even within the RPA, Eq.~(\ref{mag03}) is derived again from Eq.~(\ref{mag3}). Additionally, from Eq.~(\ref{mag4}) with $(S_i^z)^2=1/4$, one more equation
\begin{eqnarray}
\frac{1}{2}=\frac{1}{N} \sum_{\bf q}\frac{m \Gamma_{{\bf q}}}{\omega_{\bf q}}\coth \frac{\beta \omega_{\bf q}}{2} \label{mag04}
\end{eqnarray}
is derived, which cannot be obtained by using only $\{G_{ij,\eta}^{+-}, G_{ij,\eta}^{--}, G_{ij,\eta}^{z-}\}$. However, the two Eqs. (\ref{mag03}) and (\ref{mag04}) obviously contradict each other, except in the isotropic ($\Delta=1$) case. In other words, the different values of the magnetization $m$ are obtained from each equation. This fact means that the above procedure with the setting $(S_i^y)^2= (S_i^z)^2 =1/4$, which is known as a good approximation for isotropic models, is not valid for a system with an easy-plane exchange anisotropy. The same problem should arise whenever there is an anisotropy in the plane perpendicular to the direction of the magnetization (i.e., whenever the system is not rotationally invariant around the direction of the magnetization). 

\begin{figure}[t]
\begin{center}
\includegraphics[width=85mm]{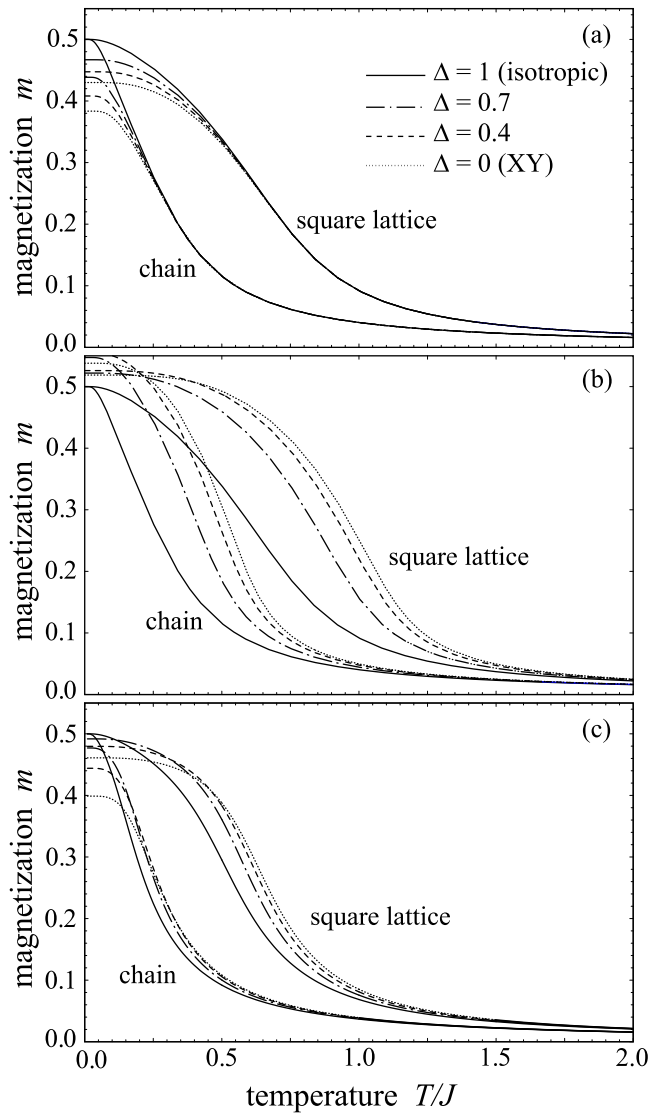}
\caption{\label{fig1}The temperature dependence of the magnetization of an easy-plane ferromagnet at $h/J=0.1$ for $\Delta=1$ (the isotropic Heisenberg model), $\Delta=0.7$, $\Delta=0.4$, and $\Delta=0$ (the $XY$ model). Comparison between the results obtained from (a) Eq.~(\ref{mag03}), (b) Eq.~(\ref{mag04}), and (c) the QMC calculations. The error bar of the QMC results are smaller than the line width. }
\end{center}
\end{figure}

This problem may be avoided by assuming that a certain restriction has to be imposed on the choice of $S_j^\nu$ in Eq.~(\ref{Green2}). This idea was introduced by Brown~\cite{Brown} for the easy-axis case. If the choice of $S_j^\nu$ is restricted to be perpendicular to the anisotropy axis, i.e., $S_j^\nu=S_j^y$, only Eq.~(\ref{mag03}) is obtained from Eq.~({\ref{mag1}}) while Eq.~(\ref{mag2}) yields the identity $0=0$. In contrast, under the restriction that $S_j^\nu=S_j^z$, only Eq.~(\ref{mag04}) is obtained from Eq.~({\ref{mag2}}). In Fig.~\ref{fig1}, the results for the temperature dependence of $m$ obtained from each case are compared with the QMC results. In the QMC calculations, where we use the continuous-time loop algorithm,\cite{ALPS,Todo,Albuquerque} $z$ direction is taken as the quantization axis. The transverse magnetization in $x$ direction is measured by using the improved estimator technique.\cite{Brower} For each temperature, measurement is performed for $6.6 \times 10^5$ Monte Carlo steps after discarding $8 \times 10^3$ steps for thermalization. The system sizes are 128 and $32 \times 32$ for the chain and the square lattice, respectively, which are large enough to produce the data in the thermodynamic limit in the resolution of Fig.~\ref{fig1}. 

We can see at once that the results obtained from Eq.~(\ref{mag4}), which are shown in Fig.~\ref{fig1}(b), violates the inequality $|\langle {\bf S}_i\rangle |\leq S$. Thus the assumption that the choice of $S_j^\nu$ is restricted to be $S_j^z$ is obviously not correct. On the other hand, as seen in Fig.~\ref{fig1}(a), the choice $S_j^\nu=S_j^y$ (or the procedure of the previous section) yields more reasonable results in the sense that $|\langle {\bf S}_i\rangle |\leq S$ is satisfied. However, in comparison with the QMC results, the approximation gets worse as the temperature increases. In particular, the intersections of the lines for the different anisotropy parameters are found in the QMC results, whereas the lines in Fig.~\ref{fig1}(a) do not cross each other.

After all, the results obtained from either choice of $S_j^\nu$ do not agree with the QMC calculations. In the first place, the reason that a restriction on the choice of $S_j^\nu$ is present is unclear. The result of the comparison between Figs.~\ref{fig1}(a) and~\ref{fig1}(c) indicates that the procedure of the previous section is inappropriate to treat anisotropic spin systems, although it has been often used for the case of systems with an easy-axis anisotropy ($\Delta>1$) and a transverse magnetic field.\cite{Frobrich,Frobrich2,Frobrich3,Frobrich4,Jensen,Wang}
\section{\label{5}Discussion and attempt to improve the method}
We now discuss why the contradiction between Eqs.~(\ref{mag03}) and (\ref{mag04}) occurs, and attempt to improve the method. The exact values of $\langle (S_i^y)^2\rangle$ and $\langle (S_i^z)^2\rangle$ are equivalent ($=1/4$) from the properties of the spin-1/2 operators. However, since we adopt the decoupling approximation, of course, $\langle (S_i^y)^2\rangle$ and $\langle (S_i^z)^2\rangle$ have errors from the exact values, respectively. As seen from Eq.~(\ref{Hamiltonian}), the parameter $\Delta$ induces the anisotropy in $yz$ plane, and the model is not symmetric under the exchange $S^y \leftrightarrow S^z$. Thus, within the RPA [Eq.~(\ref{RPA})], the errors of $\langle (S_i^y)^2\rangle$ and $\langle (S_i^z)^2\rangle$ should also be ``asymmetric'', i.e., $|\langle (S_i^y)^2\rangle_{\rm RPA}-1/4|\neq |\langle (S_i^z)^2\rangle_{\rm RPA}-1/4|$, which causes the contradiction between Eqs.~(\ref{mag3}) and (\ref{mag4}) with the setting $\langle (S_i^y)^2 \rangle=\langle(S_i^z)^2 \rangle=1/4$, namely Eqs.~(\ref{mag03}) and (\ref{mag04}). Here the subscript ``RPA'' expressly denotes the approximate values obtained within the RPA.

For general spin $S$, in addition to the constraint
\begin{eqnarray}
\mathbf{S}_i^2=(S_i^x)^2+(S_i^y)^2+(S_i^z)^2=S(S+1), 
\label{sum}
\end{eqnarray}
we have the operator identities
\begin{subequations}
\label{iden}
\begin{eqnarray}
\prod_{p=-S}^{S} (S_i^x-p)\!\!&=&\!\!0, \label{idenx}\\
\prod_{p=-S}^{S} (S_i^y-p)\!\!&=&\!\!0, \label{ideny}
\end{eqnarray}
\end{subequations}
and
\begin{equation*}
\prod_{p=-S}^{S} (S_i^z-p)=0. \label{idenz} \tag{35c}
\end{equation*}
As is mentioned above, within Green's function theory with the RPA, these severe conditions cause the contradiction between Eqs.~(\ref{mag03}) and (\ref{mag04}). Therefore we have to think about relaxing the local restrictions on the spin operators. As is clear from Eq.~(\ref{mateq2}), the calculation of $G_{ij,\eta}^{x\nu}$ is not necessary to derive Eqs.~(\ref{mag1}) and (\ref{mag2}). Thus the decoupling approximation to the time evolution of $S_i^x$ [Eq.~(\ref{dtSx})] is not actually required. Meanwhile, as for the time evolutions of $S_i^y$ and $S_i^z$ [Eqs.~(\ref{dtSy}) and (\ref{dtSz})], one needs to employ the decoupling approximation to the higher-order terms. This implies that it is inappropriate to use Eqs.~(\ref{ideny}) or (\ref{idenz}) as a conditional equation. Actually, the obtained results are unreasonable as shown in Figs.~\ref{fig1}(a) and (b). Thus, in our method, we demand that the operator identity in the direction of the magnetization [Eq.~(\ref{idenx})] is satisfied. From Eqs.~(\ref{sum}) and (\ref{idenx}), we obtain the following conditional equation: 
\begin{eqnarray}
\langle (S_i^y)^2 \rangle_{\rm RPA}+\langle (S_i^z)^2 \rangle_{\rm RPA}=\frac{1}{2}. 
\label{condition}
\end{eqnarray}
The same condition was used by Aoki in his ``self-consistent spin-wave approach.''~\cite{Aoki} Then, from Eqs.~(\ref{mag3}) and (\ref{mag4}) with the condition given by Eq.~(\ref{condition}), the self-consistent equation 
\begin{eqnarray}
\frac{1}{2}=\frac{1}{N} \sum_{\bf q}\frac{m (\tilde{\Gamma}_{{\bf q}}+\Gamma_{{\bf q}})}{2\omega_{\bf q}}\coth \frac{\beta \omega_{\bf q}}{2}\label{mag5}
\end{eqnarray}
is obtained. Obviously, when $\Delta=1$, this equation reproduces the result obtained by the conventional RPA theory for the isotropic Heisenberg model as expected. 

\begin{figure}[t]
\begin{center}
\includegraphics[width=85mm]{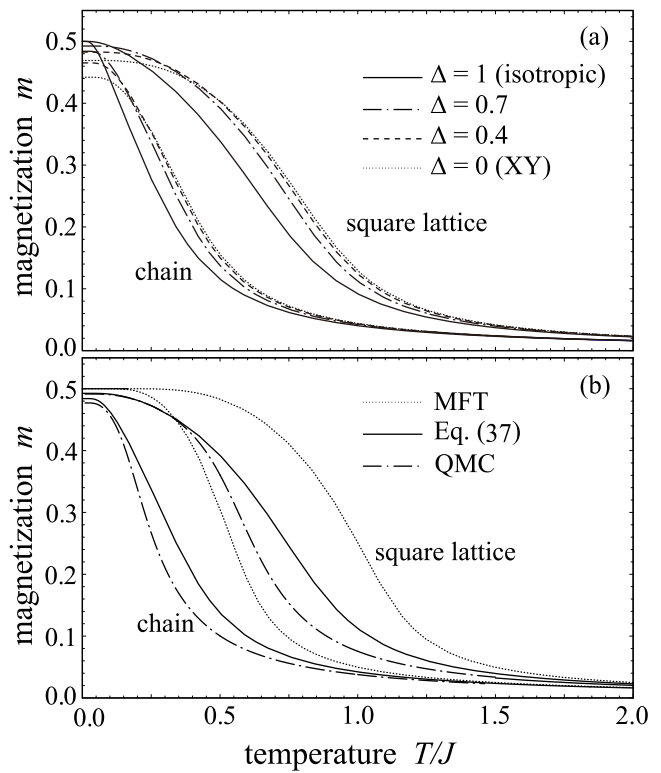}
\caption{\label{fig2}(a) The temperature dependence of the magnetization obtained from Eq.~(\ref{mag5}) at $h/J=0.1$. (b) Comparison between the results obtained from Eq.~(\ref{mag5}), the QMC calculations, and the MFT (at $h/J=0.1$ for $\Delta=0.7$). }
\end{center}
\end{figure}

As seen in Fig.~\ref{fig2}(a), 
the results from Eq.~(\ref{mag5}) are in good accord with the QMC results shown in Fig.~\ref{fig1}(c). Especially, the intersections of the lines for the different anisotropy parameters are found as expected. In Fig.~\ref{fig2}(b), we show the comparison between the results obtained from Eq.~(\ref{mag5}), the QMC calculations, and the mean-field theory (MFT). In the MFT, since the terms $S_i^yS_j^y+\Delta S_i^zS_j^z$ in Eq.~(\ref{Hamiltonian}) are neglected, the obtained magnetization curve does not depend on the anisotropy parameter: 
\begin{eqnarray}
m^{(\rm MFT)}=\frac{1}{2}\tanh \frac{\beta(h+zJm^{(\rm MFT)})}{2}.
\end{eqnarray}
Meanwhile in our procedure (and in the QMC calculations), the effect of the anisotropy can be appropriately taken into account. At low temperatures (even at $T=0$), the magnetization is suppressed by the quantum fluctuations induced by $\Delta$. In contrast, at high temperatures, the parameter $\Delta$ plays an opposite role: since the easy-plane anisotropy energetically favors the spin alignment in the easy plane, whereas the quantum fluctuations have an insignificant effect, the magnetization is enhanced as $\Delta$ decreases. As a result, the intersections of the lines for the different anisotropy parameters are found in Figs.~\ref{fig1}(c) and \ref{fig2}(a).

The application of Green's function theory to Heisenberg ferromagnets with an easy-plane anisotropy was investigated also in Refs.~\onlinecite{Rudoi3} and \onlinecite{Hu}. In particular, recently Hu $et$ $al$.~\cite{Hu} calculated the magnetization, susceptibility, and transverse correlation functions of the system described by
\begin{eqnarray}
H=-\frac{1}{2}\sum_{i,j}J_{ij}\!\left(S^x_i S^x_j\!+\!\Delta S^y_i S^y_j\!+\!S^z_i S^z_j \right)\!-\!h\!\sum_{i}S^z_i. 
\label{Hamiltonian2}
\end{eqnarray}
Obviously, this model is exactly equivalent to Eq.~(\ref{Hamiltonian}) under the rotation of the coordinate system. Using Green's functions $G_{ij,\eta}^{+-}$ and $G_{ij,\eta}^{--}$, they obtained the same results as those shown in Fig.~\ref{fig2}(a) from the condition 
\begin{eqnarray}
\langle S_i^- S_i^+ \rangle=\frac{1}{2}-\langle S_i^z \rangle. 
\label{condition2}
\end{eqnarray}
Just in the case of the coordinate system described by Eq.~(\ref{Hamiltonian2}), where the magnetization appears along $z$ axis, the condition given by Eq.~(\ref{condition2}) produced the same effect as Eq.~(\ref{condition}) of our procedure, and they obtained the same results even though they use only $G_{ij,\eta}^{+-}$ and $G_{ij,\eta}^{--}$. However, this fact does not mean that this choice of Green's functions and condition {\it always} yields reasonable results for any system. In fact, for the model in the coordinate system described by Eq.~(\ref{Hamiltonian}), the adoption of the severe conditions Eqs.~(\ref{sum_rule}) and (\ref{kinematic_interaction}) with the usual set $\{G_{ij,\eta}^{+-}, G_{ij,\eta}^{--}, G_{ij,\eta}^{z-}\}$ yields the unreasonable results, as we showed. 

\section{\label{6}SUMMARY}
In this paper, we have investigated the application of the many-body Green's function theory with the random-phase approximation to the study of easy-plane ferromagnets in an in-plane magnetic field. If there is an anisotropy in the plane perpendicular to the direction of the magnetization, then special attention is required for the choices of Green's functions and the conditions to determine the magnetization. First, we calculated the temperature dependence of the magnetization for various values of the anisotropy parameter $\Delta$ by using the usual set of Green's functions $\{G_{ij,\eta}^{+-}, G_{ij,\eta}^{--}, G_{ij,\eta}^{z-}\}$ and the conditions $\{ S_i^+,S_i^- \}=1$ and $(S_i^-)^2=0$. The obtained results did not agree with the QMC calculations on the point that no intersections of the lines for the different anisotropy parameters were found [see Figs.~\ref{fig1}(a) and \ref{fig1}(c)], which means that this procedure cannot appropriately take into account the effect of the anisotropy.

Next, by considering all combinations of two spin operators, $G_{ij,\eta}^{\mu,\nu}~(\mu,\nu=x,y,z)$, we derived the additional equation, which is not obtained by using the previous procedure. Then we showed that the two equations contradict each other under the conditions $(S_i^y)^2=(S_i^z)^2=1/4$. The same problem should also arise for higher spins, as long as the system does not have rotational symmetry around the direction of the magnetization. To avoid this contradiction within the RPA, we relaxed the restrictions; we demanded that only the operator identity in the direction of the magnetization is satisfied. For example, in the case that the magnetization appears along the $x$ axis, the condition $\langle (S_i^y)^2 \rangle_{\rm RPA}+\langle (S_i^z)^2 \rangle_{\rm RPA}=1/2$ was adopted. The results obtained by this procedure were in good agreement with the QMC calculations. This means that our method can appropriately take into account the effect of the anisotropy.

In this paper, we focused on the spin-1/2 case. Here, the extension to higher spins $S\geq 1$ will be presented. For example, according to our procedure, we have the relations ${\bf S}_i^2=2$ and $(S_i^x+1)S_i^x(S_i^x-1)=0$ for the spin-$1$ case. There is one more unknown, $\langle (S_i^x)^3 \rangle$, than $S=1/2$, and thus an additional equation is required to close the system of equations. To this end, one can adopt the following relation: 
\begin{eqnarray}
\langle \{S_i^y,S_i^x S_i^y\}\rangle+\langle\{S_i^z,S_i^x S_i^z\} \rangle=3\langle S_i^x \rangle-2\langle (S_i^x)^3 \rangle. 
\end{eqnarray}
The expression for the left-hand side is obtained from Green's functions $\langle\langle S_i^y;S_j^x S_j^\nu \rangle \rangle$ and $\langle\langle S_i^z;S_j^x S_j^\nu \rangle \rangle$ after some algebra similar to the one given in Sec.~\ref{4}. In the same way, for general spin $S$,  the system of self-consistent equations is obtained from the expressions for $\langle \{S_i^y,(S_i^x)^n S_i^y\}\rangle+\langle\{S_i^z,(S_i^x)^n S_i^z\} \rangle$ ($n=0,1,\cdots,2S-1$) and the conditions given by Eqs.~(\ref{sum}) and (\ref{idenx}).

Finally, the case of easy-axis anisotropy ($\Delta>1$) is also addressed. In this case, it is well known that the transverse magnetic field $h$ induces the spin reorientation transition for both the ferromagnetic ($J>0$) and antiferromagnetic ($J<0$) cases.\cite{Frobrich,Reinecke,Rudoi4,Brown,Frobrich2,Frobrich3,Frobrich4,Jensen,Caux,Wang,Dmitriev,Schwieger,Pini,Jensen2,Yamamoto} In order to calculate the magnetic properties, many previous authors~\cite{Frobrich,Frobrich2,Frobrich3,Frobrich4,Jensen,Wang} have employed the Green's function theory with the set of Green's functions $\{G_{ij,\eta}^{+-,mn}, G_{ij,\eta}^{--,mn}, G_{ij,\eta}^{z-,mn}\}$ and the severe conditions such as Eqs.~(\ref{sum_rule}) and (\ref{kinematic_interaction}) in the RPA. In other words, they have employed the method presented in Sec.~\ref{3} or its extended versions for higher spins. However, according to the results we showed for the easy-plane case, this choice of Green's functions and the conditions is not always appropriate. Actually, unacceptable results, especially the violation of the property of spin-$S$ systems, $|\langle {\bf S}_i\rangle |\leq S$, were found in some previous works (for example, see Ref.~\onlinecite{Frobrich4}; although it is not explicitly stated, the same problem is found also in some other works). The theoretical treatment of these systems is generally more complicated than the easy-plane cases since the transverse magnetic field causes changes in the direction of the magnetization. Nevertheless, the results reported here will be helpful for investigating these problems.

\acknowledgments
The QMC simulation in the present paper has been done by using the ALPS/looper library.\cite{ALPS,Todo,Albuquerque} One of the authors (D.Y) would like to thank K. Kamide for valuable comments and discussions. This work is supported by The 21st Century COE Program (Holistic Research and Education Center for Physics of Self-organization Systems) at Waseda University from the Ministry of Education, Culture, Sports, Science and Technology of Japan. 
\begin{appendix}
\section{Green's function theory}\label{appA}
We briefly review the many-body Green's function theory. The retarded commutator ($\eta=-1$) or anticommutator ($\eta=+1$) Green's function is defined by 
\begin{eqnarray}
G_{\eta}^{AB}(t-t')&=&\langle\langle A(t-t');B\rangle\rangle_{\eta}\nonumber\\ 
&=&-i\theta(t-t')\langle [A(t),B(t')]_{\eta} \rangle,
\label{Green}
\end{eqnarray}
where $\theta(t-t')$ denotes the step function and $\langle \cdots \rangle={\rm Tr}(e^{-\beta H}\cdots)/{\rm Tr}(e^{-\beta H})$ is an average over the ensemble with the inverse temperature $\beta=1/T$. $[A,B]_{-1}=[A,B]$ and $[A,B]_{+1}=\{A,B\}$ denote the commutator and anticommutator of operators $A$ and $B$, respectively. Green's function satisfies the following equation of motion in energy space: 
\begin{eqnarray}
\omega G_{\eta}^{AB}(\omega)=\langle [A, B]_{\eta}\rangle+\langle\langle[A,H];B\rangle\rangle_{\eta,\omega}.
\label{Eqofmotion0}
\end{eqnarray}
To obtain the solution of $G_{\eta}^{AB}(\omega)$, a certain decoupling approximation of higher-order Green's functions in the right-hand side of the above equation is usually adopted. Then the expectation values and correlation functions can be calculated from the spectral theorem~\cite{Tyablikov,Frobrich}
\begin{eqnarray}
\langle AB\rangle\!\!&=&\!\!\frac{i}{2\pi} \!\lim_{\delta\to 0}\int_{-\infty}^{\infty} \!\!\!d\omega \frac{G_{-1}^{AB}(\omega+i\delta)\!-\!G_{-1}^{AB}(\omega-i\delta)}{1-e^{-\beta\omega}}\nonumber\\
&&\!\!+\frac{C_{+1}^{AB}}{2},\label{spectral1}
\end{eqnarray}
and
\begin{eqnarray}
\langle BA\rangle\!\!&=&\!\!\frac{i}{2\pi} \!\lim_{\delta\to 0}\int_{-\infty}^{\infty} \!\!\!d\omega \frac{G_{-1}^{AB}(\omega+i\delta)\!-\!G_{-1}^{AB}(\omega-i\delta)}{e^{\beta\omega}-1}\nonumber\\
&&\!\!+\frac{C_{+1}^{AB}}{2},
\label{spectral2}
\end{eqnarray}
where 
\begin{eqnarray}
C_{\eta}^{AB}=\lim_{\omega\rightarrow 0}\omega G_{\eta}^{AB}. 
\end{eqnarray}
If the anticommutator Green's function has a first-order pole at $\omega=0$, the terms $C_{+1}^{AB}/2$ in Eqs.~(\ref{spectral1}) and (\ref{spectral2}) are required. Meanwhile, it is well known that the commutator Green's function has no pole at $\omega=0$ in any case. Incidentally, Eqs.~(\ref{spectral1}) and (\ref{spectral2}) can be rewritten in the following more convenient form:
\begin{eqnarray}
\langle \{A,B\}\rangle\!\!&=&\!\!\frac{i}{2\pi} \!\lim_{\delta\to 0}\int_{-\infty}^{\infty} \!\!\!d\omega \left[ G_{-1}^{AB}(\omega+i\delta)\right.\nonumber\\
&&\!\!\left.-G_{-1}^{AB}(\omega-i\delta)\right]\coth \frac{\beta \omega}{2}+C_{+1}^{AB},\label{spectral3}
\end{eqnarray}
and
\begin{eqnarray}
\langle[A,B]\rangle\!\!&=&\!\!\frac{i}{2\pi} \!\lim_{\delta\to 0}\int_{-\infty}^{\infty} \!\!\!d\omega  \nonumber \\
&&\!\!\times \left[G_{-1}^{AB}(\omega+i\delta)\!-\!G_{-1}^{AB}(\omega-i\delta)\right].
\label{spectral4}
\end{eqnarray}
In calculations of spin systems, Eq.~(\ref{spectral3}) generally plays the role of the self-consistent equation, whereas Eq.~(\ref{spectral4}) is trivial and gives no informations.

\end{appendix}
\newpage 

%

\end{document}